\documentclass[aps,prm,amsmath,amssymb,reprint,superscriptaddress,]{revtex4-2}
\usepackage{charter,graphicx,verbatim,threeparttable,float,amssymb,gensymb }
\newcommand{\AVS}{\textit{A}V$_3$Sb$_5$}
\newcommand{\KVS}{KV$_3$Sb$_5$}
\newcommand{\RVS}{RbV$_3$Sb$_5$}
\newcommand{\CVS}{CsV$_3$Sb$_5$}
\newcommand{\dtcdw}{$\Delta T_\text{CDW}$}
\newcommand{\tcdwlin}{$T_\text{CDW,Linear}$}
\newcommand{\tcdw}{$T_\text{CDW}$}
\newcommand{\tc}{$T_\text{c}$}
\begin{document}

\preprint{APS/123-QED}

\title{Complete miscibility amongst 
 the \textit{A}V$_3$Sb$_5$ kagome superconductors: design of mixed \textit{A}-site \textit{A}V$_3$Sb$_5$ (\textit{A}: K, Rb, Cs) alloys}

\author{Brenden R. Ortiz} \email{ortiz.brendenr@gmail.com}
 \affiliation{Materials Department, University of California Santa Barbara, Santa Barbara, CA 93106, USA}

 \author{Andrea N. Capa Salinas}
 \affiliation{Materials Department, University of California Santa Barbara, Santa Barbara, CA 93106, USA}

\author{Miles J. Knudtson}
 \affiliation{Materials Department, University of California Santa Barbara, Santa Barbara, CA 93106, USA}
 
\author{Paul M. Sarte}
 \affiliation{Materials Department, University of California Santa Barbara, Santa Barbara, CA 93106, USA}
 
 \author{Ganesh Pokahrel}
 \affiliation{Materials Department, University of California Santa Barbara, Santa Barbara, CA 93106, USA}
 
 \author{Stephen D. Wilson} \email{stephendwilson@ucsb.edu}
 \affiliation{Materials Department, University of California Santa Barbara, Santa Barbara, CA 93106, USA}

\date{\today}

\begin{abstract}
 In this work we explore the chemical-property phase diagram of the \AVS~family through \textit{A}-site alloying.  We demonstrate full miscibility of the alkali-site, highlighting that the three parent compounds are the terminal ends of a single solid-solution. Using both polycrystalline and single crystal methods, we map the dependence of the two primary electronic instabilities: (1) the onset of charge density wave (CDW) order (\tcdw) and (2) the onset of superconductivity (\tc) with alkali-site composition. We show continuous trends in both \tcdw~and \tc, including a region of enhanced CDW stability in K$_{1-x}$Cs$_{x}$V$_3$Sb$_5$ alloys. Together, our results open new routes for chemical perturbation and exploration of the chemical-property relationships in the class of \AVS~kagome superconductors.
\end{abstract}

\maketitle

\section{Introduction}

\begin{figure}
	\includegraphics[width=\linewidth]{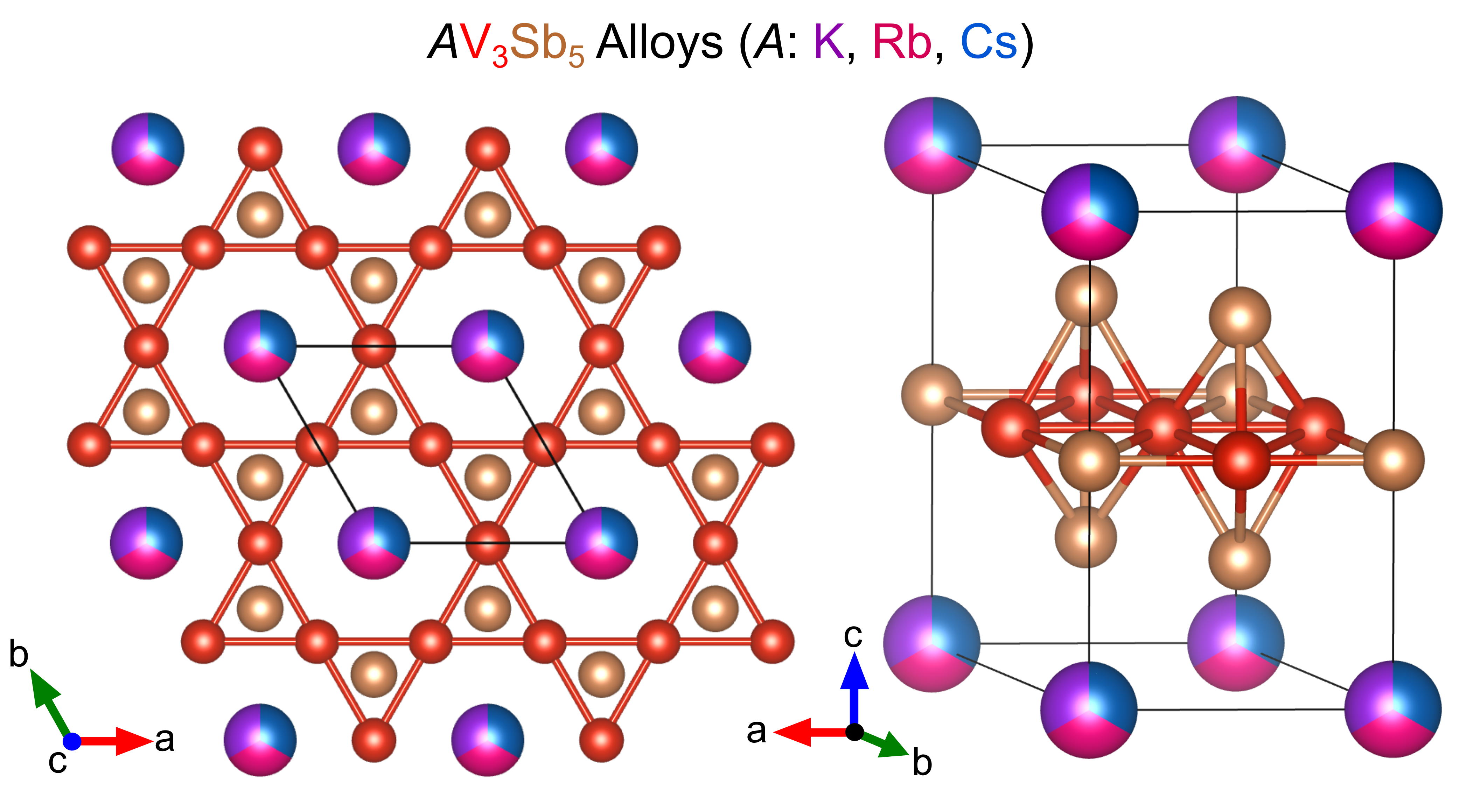}
	\caption{The \AVS~(\textit{A}: K, Rb, Cs) kagome superconductors are a family of layered, exfoliable, kagome metals consisting of a structurally perfect lattice of vanadium at room temperature. All three compounds exhibit a charge density wave transition and a lower temperature superconducting ground state. The crystal structure is illustrated above with the alloyed \textit{A}-site highlighted.}
	\label{fig:1}
\end{figure}

The recent discovery of the \AVS~(\textit{A}: K, Rb, Cs) kagome superconductors \cite{ortiz2019new,ortizCsV3Sb5} (Figure \ref{fig:1}) highlights some of the unique possibilities present in  kagome metals. The interplay between the emergence of a charge density wave state ($T_\text{CDW}$: 84\,K, 104\,K, 94\,K) \cite{ortiz2020KV3Sb5,RbV3Sb5SC,ortizCsV3Sb5} and superconductivity ($T_\text{c}$: 1.0\,K, 0.8\,K, 2.5\,K) \cite{ortiz2020KV3Sb5,RbV3Sb5SC,ortizCsV3Sb5} in these compounds has generated considerable research into the \AVS~family. Underlying this interest are hints of exotic properties ranging from pair density wave superconductivity \cite{chen2021roton,ge2022discovery}, orbital magnetism (chiral charge density wave) \cite{xu2022universal,mielke2022time,yu2021evidence,guo2022switchable,jiang2021unconventional}, and topological surface states \cite{hu2022topological,kang2022twofold,hu2022rich} present in the superconducting phase.

Due in part to its ease of growth, relatively higher \tc, and higher quality crystals, CsV$_3$Sb$_5$ has been the primary focus of the \AVS~family. Curiously, the current literature hints that many features of  CsV$_3$Sb$_5$'s charge density wave state are unique to this variant. In the \AVS~family, the real component of the CDW manifests as a 2$\times$2 in-plane superlattice. The associated structural distortion results in individual kagome planes adopting either the ``star-of-David (SoD)'' or ``tri-hexagonal (TrH)'' motif \cite{PhysRevLett.127.046401,PhysRevLett.127.217601,PhysRevB.104.214513}. Unlike the K- and Rb-based compounds, CsV$_3$Sb$_5$ manifests a combination of both distortion types within a mixed CDW state with both 2$\times$2$\times$4 and 2$\times$2$\times$2 order possible \cite{ortiz2021fermi,hu2022coexistence,kang2022microscopic}.  Furthermore, the emergence of multiple different temperature regimes within the CDW state, and an apparent sensitivity to thermal history (heating rates, hysteresis, quenching) are thus far unique to CsV$_3$Sb$_5$ \cite{stahl2022temperature,xiao2022coexistence}. Small quantities of dopants have also driven changes in the dimensionality of the CDW in CsV$_3$Sb$_5$, suggesting a complex interplay of coexisting or competing CDW instabilities \cite{kautzsch2022incommensurate}. Differences amongst the other compounds have been noted as well, with KV$_3$Sb$_5$ being the only compound to apparently lack the 4$a_0$ unidirectional stripe charge order observed at the surface in scanning tunneling microscopy studies \cite{zhao2021cascade,li2022emergence,yu2022evolution}. 

Of particular note are the various high-pressure and extrinsic doping studies (e.g. Sn\cite{oey2022Cs,oey2022KRb}, Ti\cite{yang2022titanium,liu2021dopingTi}, Nb\cite{li2022tuning}, Ta\cite{liu2022evolution}, Mo\cite{liu2022evolution}, and O\cite{song2021competition}), which have highlighted additional differences between the electronic phase diagram of CsV$_3$Sb$_5$ and the Rb/K-based variants. CsV$_3$Sb$_5$ exhibits a "double-dome" feature in both high-pressure ($<$10\,GPa) and chemical-doping studies, whereas the Rb- and K- variants appear to show only ``single-dome'' superconductivity when perturbed \cite{oey2022Cs,oey2022KRb,du2021pressure,chen2021double,zhu2022double,yu2021unusual}. These differences support the notion of a unique starting CDW state in CsV$_3$Sb$_5$, an effect likely driven by subtle structural (e.g. interlayer distances) differences between the alkali metal planes and the Zintl-like V-Sb networks. 

Chemical alloying is a powerful tool for exploring the relationship between alkali-site character and CDW properties in the \AVS~system.  Alloys between the three variants stand to provide a tool for exploring a crossover between CDW states, helping to elucidate the underlying mechanism. Here we report the formation and properties of the full suite of isoelectronic alloys (K,Rb,Cs)V$_3$Sb$_5$. We demonstrate full miscibility of the alkali site, highlighting that the entire \AVS~family shares a single solid-solution. A continuous, though nonlinear, dependence of the CDW transition on alkali-site character is shown using 45 unique polycrystalline samples, and a potential crossover along the line of K$_{1-x}$Cs$_{x}$V$_3$Sb$_5$ alloys is noted. Furthermore, we present the means to design and grow single crystal, mixed \textit{A}-site \textit{A}V$_3$Sb$_5$ alloys, highlighting changes in $T_c$ as well. This work expands the chemical phase space of the \AVS~kagome superconductors, and demonstrates a framework wherein \KVS, \RVS, and \CVS~are viewed as terminal end points of a single phase diagram.

\section{Experimental Methods}

\subsection{Polycrystalline Synthesis}

To generate the large number of polycrystalline \AVS~samples desired in this study, a method similar to that published previously was used \cite{ortiz2019towards}, appropriately modified for use under inert atmosphere. Amorphous precursor feedstocks with stoichiometric compositions of A$_{1.05}$V$_3$Sb$_5$ were created for each of the three parent compounds \KVS, \RVS, and \CVS. Elemental K ingot (Alfa, 99.8~\%), Rb ingot (Alfa, 99.75~\%), Cs liquid (Alfa, 99.98~\%), V powder (Sigma, 99.9~\%), and Sb shot (Alfa, 99.999~\%) were sealed into pre-seasoned tungsten carbide ballmill vials under argon and milled for 90\,min in a SPEX 8000D dual mixer/mill. As-received vanadium powder was first purified using a combination of EtOH and concentrated HCl to remove residual oxides on the powder. Resulting feedstocks were then ground and passed through a 100\,$\mu$m sieve and stored. 

The three feedstock powders were subsequently combined in various ratios to form amorphous \AVS~alloy precursors. The mixtures were sealed into 5\,mL (SPEX 3127) vials. We subdivided the \KVS--\RVS--\CVS~phase diagram into 1/8 increments, for a total of 45 unique polycrystalline samples. Vials were sealed under argon and milled for 60\,min in batches of 8, utilizing the parallel nature of the SPEX 3127 adapter. The individual (amorphous) mixtures of (K,Rb,Cs)V$_3$Sb$_5$ were then extracted, ground, and passed through 100\,$\mu$m sieves. The powders were sealed in fused silica ampoules under 1\,atm of argon and batch annealed at 550\degree\,C for 48\,h. Transformed polycrystalline powders were extracted from the ampoules under argon and stored.

Testing for other isoelectronic alloy solubility limits with \CVS~was performed in an analogous way. Feedstocks of hypothetical mixtures corresponding to CsNb$_3$Sb$_5$, CsTa$_3$Sb$_5$, NaV$_3$Sb$_5$, CsV$_3$As$_5$, CsV$_3$Bi$_5$, and TlV$_3$Bi$_5$ were generated using Nb powder (Alfa, 99.8~\%), Ta powder (Alfa, 99.97~\%), Na ingot (99.8~\%), As chunk (Alfa, 99.9999~\%), and Bi rod (Alfa, 99.99~\%). Alloys were probed by making stoichiometric mixtures of the amorphous precursor phases with the parent \CVS~precursor phase. We also tested mixtures to confirm that alternate \textit{AM}$_3$\textit{X}$_5$ phases \textit{do not} form.

\subsection{Single Crystal Synthesis}

Single crystal \AVS~alloys were grown by a modified self-flux method, as performed in prior studies \cite{ortiz2020KV3Sb5,ortizCsV3Sb5,ortiz2021fermi}. Analogous to the polycrystalline study, we synthesized large feedstocks of the parent flux precursors and subsequently recombined them to rapidly generate a series of alloyed fluxes. The overall flux composition is slightly modified from prior reports \cite{ortiz2021fermi}, and corresponds to \textit{A}$_{20}$V$_{15}$Sb$_{120}$. Flux feedstocks were produced using the same reagents as the polycrystalline synthesis. The constituent elements were sealed into pre-seasoned tungsten carbide ballmill vials under argon and milled for 90\,min in a SPEX 8000D dual mixer/mill. 

The three parent feedstocks: K$_{20}$V$_{15}$Sb$_{120}$, Rb$_{20}$V$_{15}$Sb$_{120}$, and Cs$_{20}$V$_{15}$Sb$_{120}$ were then blended to produce the desired (K,Rb,Cs)$_{20}$V$_{15}$Sb$_{120}$ flux precursor. Powders were loaded into 2\,mL high-density alumina (Coorstek) crucibles and loaded into steel tubes under 1\,atm of argon. Samples were heated to 1000\degree\,C at 200\degree\,C/hr and subsequently soaked at 1000\degree\,C for 12\,h. Samples were then cooled relatively quickly to 900\degree\,C at 5\degree\,C/hr and then slow cooled to 600\degree\,C at 2\degree\,C/hr. Fluxes were cooled completely and crystals were extracted mechanically. The resulting crystals are thin hexagonal flakes with a metallic silver luster. Typical linear dimensions are 2\,mm--1\,cm depending on the composition of choice. Crystals are air-stable and readily exfoliable.

\subsection{Structural, Chemical, and Electronic Properties Characterization}

Polycrystalline sample purity and crystallinity was examined with powder x-ray diffraction (XRD) measurements at room temperature on a Panalytical Empyrean diffractometer (Cu K$_{\alpha_{1,2}}$) in Bragg-Brentano ($\theta$-$\theta$) geometry. Rietveld refinement of powder XRD patterns was performed using \texttt{TOPAS Academic} v6 \cite{Coelho}. Structural models and visualization utilized the \textsc{VESTA} software package \cite{Momma2011}. Elemental compositions of both powders and single crystal samples were measured using a Hitachi TM4000 Plus scanning electron microscope (SEM) where energy-dispersive spectroscopy (EDS) measurements were performed under an accelerating voltage of 20\,kV. Single crystal samples were exfoliated before measurement. 

\begin{figure*}
	\includegraphics[width=1\textwidth]{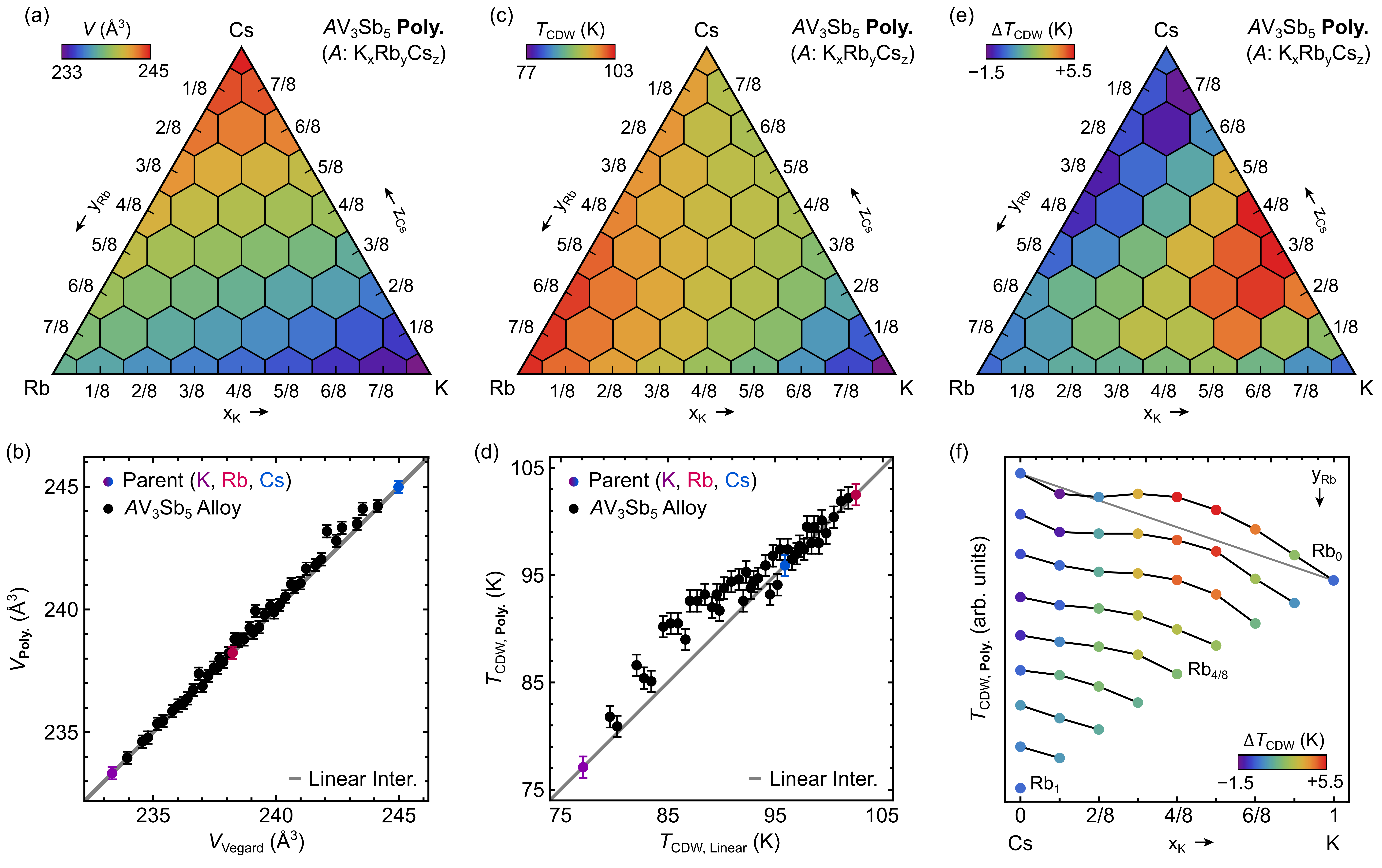}
	\caption{Experimental synthesis of \AVS~polycrystalline alloys (a,b) reveals a full solid-solution between the three parent compounds that obeys linear interpolation (e.g. Vegard's Law). The charge density wave temperature \tcdw~also trends smoothly between the termini (c), though the trends are not strictly linear (d). A plot of \dtcdw~as a function of composition reveals a pocket of enhanced \tcdw~along the K$_{1-x}$Cs$_{x}$V$_3$Sb$_5$ alloy line which is smooth and continuous (e). Note that the curves in (f) are offset for visual clarity, though the \textit{scales} of the curves are consistent.}
	\label{fig:2}
\end{figure*}

Magnetization measurements determining the CDW state in \AVS~ samples were performed using a 7~T Quantum Design Magnetic Property Measurement System (MPMS3) SQUID magnetometer in vibrating-sample magnetometry (VSM) mode. Powders were placed in polypropylene capsules and mounted in a brass sample holder. Single crystal samples were mounted to quartz holders using GE varnish. Measurements were collected at 1\,T under zero-field-cooled (ZFC) conditions.

Electrical resistivity measurements of the superconducting state in single crystal samples were performed using a Quantum Design 14~T Dynacool Physical Property Measurement Systems (PPMS) equipped with a dilution refrigerator (DR) insert and the electronic transport option (ETO). Single crystals were mounted to the sample stage using a small quantity of cigarette paper and GE varnish to ensure electrical isolation and thermal contact. Samples were then exfoliated and contacts estabilished using silver paint (DuPont cp4929N-100) and gold wire (Alfa, 0.05mm premion 99.995\%). We used a probe current of 1\,mA and frequency $\approx$100\,Hz to measure the onset of superconductivity under zero-field conditions.

\section{Polycrystalline Results \& Discussion}

Polycrystalline samples can be made quickly and with precise control over stoichiometry, providing an efficient means of traversing potential alloy spaces. While this manuscript will explicitly focus on alloys between \KVS, \RVS, and \CVS, we note here that we also performed a cursory exploration of other isoelectronic alloys with \CVS. We explored Bi$_\text{Sb}$, As$_\text{Sb}$, Nb$_\text{V}$, Ta$_\text{V}$, and Na$_\text{Cs}$. A summary of the solubility limits as determined by diffraction is shown in Table \ref{tab:1}. None of the above produced new \textit{AM}$_3$\textit{X}$_5$ compounds, and attempts to alloy the individual components within \CVS~ resulted in relatively low solubility. This was somewhat unexpected, considering that some electronically active dopants have exhibited extensive solubility (e.g. Sn$_\text{Sb}$ reaches nearly CsV$_3$Sb$_4$Sn) \cite{oey2022Cs}. We explored other isoelectronic combinations with Tl$^{+}$, Au$^{+}$, and Hg$^{+}$, though no new variants emerged.

\begin{table}[t]
\caption{Approximate solubility limits in polycrystalline samples of \textit{AM}$_3$\textit{X}$_5$--CsV$_3$Sb$_5$ obtained from diffraction measurements.}

    \renewcommand{\tabcolsep}{9pt}
	\begin{tabular}{c|ccc}
		\hline \hline
		Element & Site & Formula & Solubility Limit \\ \hline
		Na & Na$_\text{Cs}$ &  Na$_{x}$Cs$_{1-x}$V$_3$Sb$_5$ & x:[0,0.2]           \\
		K &  K$_\text{Cs}$ &  K$_{x}$Cs$_{1-x}$V$_3$Sb$_5$ & x:[0,1.0]             \\
		Rb  &  Rb$_\text{Cs}$ &  Rb$_{x}$Cs$_{1-x}$V$_3$Sb$_5$ & x:[0,1.0]  \\
		As  &  As$_\text{Sb}$ & CsV$_3$Sb$_{5-x}$As$_x$ & x:[0,0.3]    \\
		Bi  &  Bi$_\text{Sb}$ & CsV$_3$Sb$_{5-x}$Bi$_x$ & x:[0,0.4]   \\
		Nb  &  Nb$_\text{V}$ & CsV$_{3-x}$Nb$_{x}$Sb$_5$ & x:[0:0.2]   \\
		Ta  &  Ta$_\text{V}$ & CsV$_{3-x}$Ta$_{x}$Sb$_5$  & x:[0:0.1]  \\
		\hline \hline
		\end{tabular}
		\label{tab:1}
\end{table}

Now turning to the synthesis and characterization of \AVS~alloys within the \KVS--\RVS--\CVS~ phase space, Figures \ref{fig:2}(a,b) shows the resulting cell volume of the \AVS~structure as a function of composition throughout the entire \KVS--\RVS--\CVS~phase diagram.  Figure \ref{fig:2}(a) clearly highlights the continuous change in cell volume between the three end-members, as would be expected for a full-solid solution.  In Figure \ref{fig:2}(b), an alternate depiction of Figure \ref{fig:2}(a) is shown where the $x$-axis is recast as the \textit{theoretical cell volume} under the auspice of Vegard's Law (linear interpolation of the lattice parameters of pure \KVS, \RVS, and \CVS~powders). The parent compositions have been highlighted for reference, and a 1:1 correspondence is drawn as a gray line. The close agreement of the volume in Figure \ref{fig:2}(b) with the linear projection indicates excellent agreement with Vegard's Law, and confirms that the entire \AVS~family exists as a single solid-solution.

It is worth highlighting the consequence of the \AVS~single solid-solution. This indicates that \KVS, \RVS, and \CVS~are not line-compounds, but terminal ends of a full solid-solution in a three-dimensional phase space. The continuous chemical alloying also presents an opportunity to investigate chemical-property changes throughout the \AVS~family. Properties attributed as unique to \CVS~may exhibit interesting ``crossover'' points as we alloy with other endpoints. For example, as a potential future study, the incremental addition of Rb or K into \CVS~may help illuminate the nature of the 2$\times$2$\times$4 CDW order which is only reported in \CVS~ \cite{ortiz2021fermi,hu2022coexistence,kang2022microscopic}. 

Having established that a solid-solution is possible, we next turn to evolution of the CDW state across the series of \AVS~alloy powders. Figure \ref{fig:2}(c) shows a heat map of the CDW temperature \tcdw~as a function of alloying composition. For a consistent, unbiased extraction of \tcdw~from the magnetization data, \tcdw~is defined as the peak in the derivative $d(MT)/dT$. A clear continuous gradient between the three parent compounds is observed, though Figure \ref{fig:2}(c) disguises some nuances in \tcdw~as a function of composition. Figure \ref{fig:2}(d) shows the \tcdw~with the x-axis again recast as the \textit{theoretical} \tcdw~if we presume a simple linear interpolation between the three parent compounds. Again we highlight the three, pure parent compositions on the graph. The data near \RVS~agrees well with a linear interpretation; however substantial deviations from nominal values are revealed as the transition towards \KVS~is approached. 

Further exploring this departure, Figure \ref{fig:2}(e) transforms the data from Figure \ref{fig:2}(c,d) as \dtcdw, where \dtcdw~=\tcdw$-$\tcdwlin. This allows the deviation from linearity in \tcdw~to be visualized as a function of composition. The most prominent feature occurs along the K$_{1-x}$Cs$_{x}$V$_3$Sb$_5$ series of alloys, where we observe a substantial enhancement of the CDW above the nominal (linearly interpolated) value. Figure \ref{fig:2}(f) plots 1D slices through the \dtcdw~data, allowing an alternate perspective. Each curve connected by black is a iso-Rb composition, demonstrating the change in \tcdw~as a function of K:Cs ratio. The gray line on the Rb$_0$ contour is the result from linear interpolation, and the color scheme is true to \dtcdw. The iso-Rb contours are offset graphically in the y-direction for visual clarity. Interestingly, if one was to continuously grade a composition from \CVS--\RVS, and then from \RVS--\KVS, the trends in \tcdw~are nearly linear, suggesting that something unique may happen between K:Cs mixtures where a pocket of ``enhanced'' \tcdw~can be observed along the K$_{1-x}$Cs$_{x}$V$_3$Sb$_5$ line.

We now turn to analysis of how the superconducting transition evolves across the phase diagram.  In prior studies, polycrystalline samples and single crystals have always shown reasonable agreement with onset temperature of the CDW state in \AVS. However, the superconducting transition can be seemingly degraded by disorder/strain effects native to powders. For instance, powder samples require an additional annealing step to obtain fully superconducting volume fractions \cite{ortizCsV3Sb5,oey2022Cs,oey2022KRb}. This is likely due to powders exhibiting an enhanced sensitivity to strain and deformation, where even slight grinding in an agate mortar and pestle is sufficient to suppress the superconducting state. Prior to annealing, such powders remain highly crystalline by XRD and the CDW is unaffected -- making powders non-ideal for measurement of the superconductivity.  As a result, we also created a smaller set of single crystal samples to explore the evolution of superconductivity in tandem to the CDW state across the \AVS~alloy space.

\begin{figure*}
	\includegraphics[width=1\textwidth]{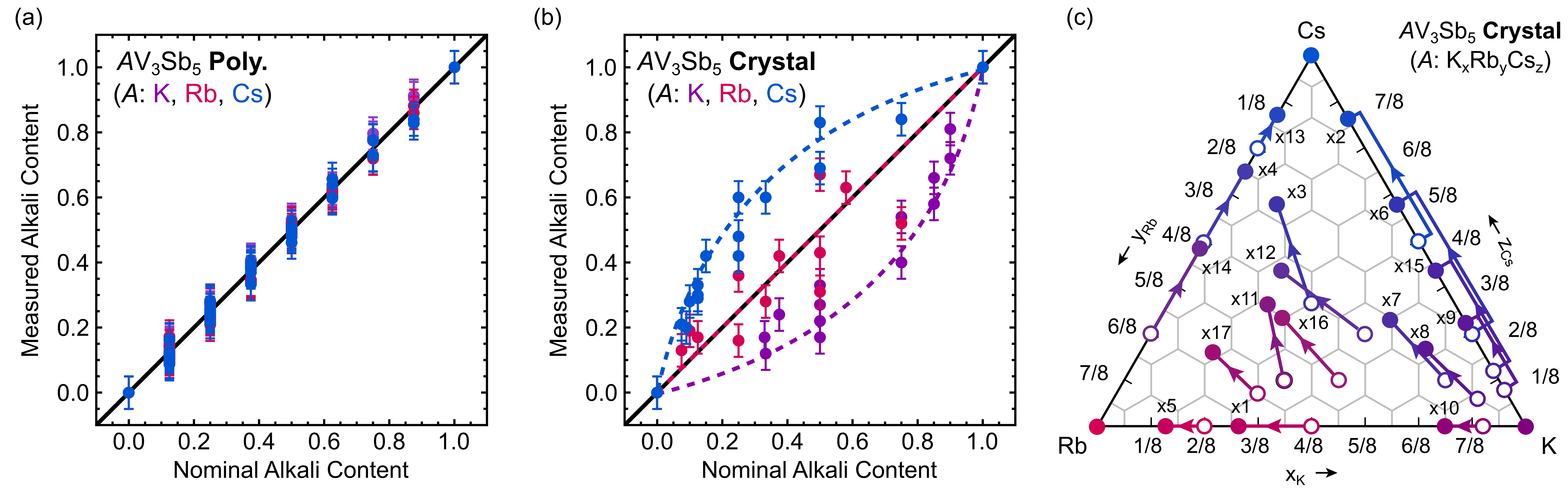}
	\caption{Here we compare compositional control in polycrystalline and single-crystal samples of \AVS~alloys. Energy dispersive spectroscopy (EDS) on polycrystalline samples (a) highlights that all samples exhibit alkali-metal ratios consistent with their nominal loading. Single crystal samples (b), however, exhibit alkali metal content that deviates from their nominal values. We observe that (K, Rb, Cs) mixtures tend to be deficient in K and exhibit excess Cs. These relationships can be visualized as a vector map (c) in the \KVS--\RVS--\CVS~ space, highlighting the ``loss vector.'' While empirical, this provides a means to target a given \AVS~composition.}
	\label{fig:3}
\end{figure*}

\section{Single Crystal Results \& Discussion}

A series of 20 single crystals of varying target compositions were grown, and Figure \ref{fig:3} compares the resulting compositional control in polycrystalline samples versus single crystals. Figure \ref{fig:3}(a) shows the experimentally measured alkali content in the 45 polycrystalline samples shown in Figure \ref{fig:1} as a function of their \textit{nominal} composition. All powder samples fall within the standard error of EDS chemical composition measurements. However, flux-grown crystals of \AVS~show substantially more complex behavior. Figure \ref{fig:3}(b) shows the compositional data for the 20 single crystal samples (3 parents, 17 alloys). \textit{Relative to their nominal compositions}, samples that include K are always deficient in K. The opposite is observed for Cs, which is always more abundant than the nominal content. Interestingly, the \textit{ratio} of Rb to the other elements appears to be largely consistent with the nominal, target composition.

Figure \ref{fig:3}(c) provides an alternate way to visualize compositional trends in the flux grown crystals. Open circles highlight starting ``nominal'' compositions, while the closed circles indicate the final ``measured'' composition. The compositional vector is referred to as the ``loss vector'' and can be used to map nominal alkali loading onto the expected composition. Several qualitative rules can be inferred: (1) In (K,Rb,Cs) mixtures, the relative abundance trends towards the Rb:Cs edge of the diagram, (2) In pairwise mixtures with Cs, Cs will always exist in relative abundance relative to the target composition, and (3) In pairwise mixtures, K will always exist in relative scarcity relative to the target composition. 

The difference between the nominal and measured compositions are relatively large for many compositions. Speculating, we suspect that mixtures of the three \AVS~compounds likely modify the melting point of the fluxes substantially and change the relative stability and constituent chemical potentials of the various alkali metals. However, the ``loss vectors'' shown in Figure \ref{fig:3}(c) are relatively consistent in length and direction, allowing specific compositions to be targeted empirically, even without more intimate knowledge of the liquidus surface and melting points. We note here that \textit{within} a given \AVS~alloy batch, a random selection of $>$5 crystals all exhibit consistent compositions. This implies that a given flux composition yields reproducible and homogeneous conditions within the growth. 

\begin{figure*}
\includegraphics[width=1\textwidth]{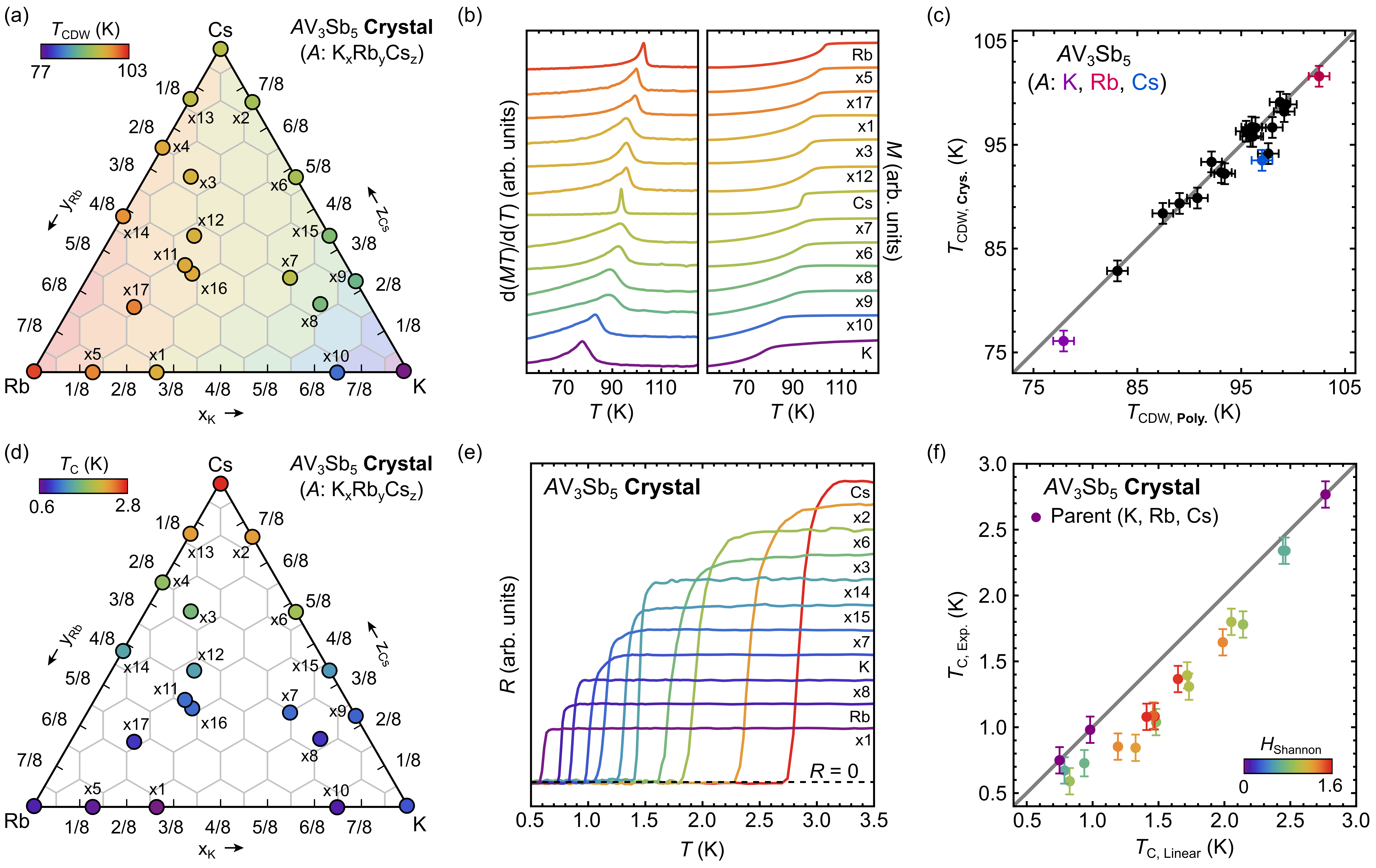}
\caption{Here we examine the charge density wave and superconducting transitions in 20 unique single crystal \AVS~samples (17 alloys). The \tcdw~(a) data extracted using the peak in the $d(MT)/dT$ data (b) is shown superimposed with the polycrystalline data. Note that $d(MT)/dT$ data and $M(T)$ is shown in arb. units and has been scaled and offset for visual comparison. Excellent agreement between the polycrystalline (interpolated experimental \tcdw~surface) and single crystal \tcdw~data (c) indicates that the crystals exhibit the same trends -- and that the polycrystalline data can serve as a good proxy for selecting compositions. Analogous superconducting transitions (d) are extracted using the zero-resistivity condition in transport data (e). Note that resistivity curves are shown in arb. units and have been scaled for visual clarity. In all cases we observe a suppression of \tc~relative to a naive, linear interpolation (f), consistent with the effect of chemical disorder.}
\label{fig:4}
\end{figure*}

The single crystal samples presented in Figure \ref{fig:3} provide another platform to explore the connection between the \textit{A}-site chemistry and the CDW state. Figure \ref{fig:4}(a) shows \tcdw~as a function of composition, again extracted using the peak in $d(MT)/dT$. The powder data has been included as a transparent background to show qualitative agreement with the prior polycrystalline data. A representative selection of the $d(MT)/dT$ curves and the raw magnetization data are also shown in Figure \ref{fig:4}(b) for comparison. For clarity we have included all single crystal magnetization results in the supplementary information \cite{ESI}. Alloyed single crystals tend to exhibit broader \tcdw~transitions, though not substantially broader than the transition in pure \KVS.

Recall that one of the unique features of the polycrystalline \tcdw~data in Figure \ref{fig:2} was the dome of enhanced \tcdw~relative to the linear interpolation. Due to the limitations of single-crystal synthesis (largely a throughput issue), the single crystal data in Figure \ref{fig:4}(a) can be considered ``aliased.'' Figure \ref{fig:4}(c) harmonizes the single crystal and polycrystalline data to determine whether we expect single crystal alloys along K$_{1-x}$Cs$_{x}$V$_3$Sb$_5$ to exhibit analogous trends. Here we show the experimentally extracted \tcdw~for single crystals against the \textit{interpolated} $T_\text{CDW,Poly}$ for powders. Note that the \textit{interpolated}  $T_\text{CDW,Poly}$ is not a linear interpolation, but a learned interpolation based on the experimental surface observed in Figure \ref{fig:2}(c-f). The 1:1 agreement in Figure \ref{fig:4}(c) highlights that the single crystal data is in excellent agreement with the polycrystalline data. Thus, the polycrystalline data can serve as a proxy to identify interesting compositions, which can be subsequently targeted using the vector map in Figure \ref{fig:3}(c).  As one exception to this, the points exhibiting largest deviation from the 1:1, crystal:powder correspondence are pure \CVS~and a Cs-rich crystal. Deviations between the CDW temperatures in powders and single crystals of pure \CVS~were reported earlier \cite{ortizCsV3Sb5, oey2022Cs}, and the origin for this difference remains an open question.

Next we examine the effect of \textit{A}-site alloying on the superconducting transitions of \AVS~alloys. Figure \ref{fig:4}(d) shows a heat map of the single crystal samples and their respective \tc, extracted from resistivity measurements at low temperatures. Figure \ref{fig:4}(e) shows a selection of the raw resistivity curves for comparison. As with the magnetization results, all resistivity results on single crystals have been included in the supplementary information \cite{ESI}. As a conservative estimate, we use the zero-resistivity condition to define \tc. In doing so, the \tc's of the parent compounds agree with published heat-capacity and Meissner state data within 0.2\,K. Note that the initial report of superconductivity in \RVS~lists \tc~as 0.92\,K \cite{RbV3Sb5SC}, though the zero-resistivity condition in that manuscript is 0.75\,K, in agreement with our current data.

Figure \ref{fig:4}(f) compares \tc~for the \AVS~alloys with a simple linear interpolation between the three parent compounds. All alloys exhibit \tc~values that are suppressed relative to the parent \KVS, \RVS, and \CVS~compounds. All samples also maintain a clear superconducting transition and zero resistivity state. To verify the bulk nature of the transition in the alloyed compositions, we performed heat capacity measurements on a heavily alloyed single crystal \cite{ESI}, confirming that the bulk nature of the superconducting state. The global suppression of \tc~is consistent with the influence of increased chemical and site disorder induced by \textit{A}-site alloying.  As means of further visualizing this, the data points in Figure \ref{fig:4}(f) are color coded using the Shannon entropy, calculated as $H = -\sum_{i=3}^n p(x_i)\log_b p(x_i)$. Here the probabilities $p(x_i)$ are approximated as the relative concentration of the three alkali-metals measured in each crystal. As a general trend, samples with higher entropy mixtures (e.g. closest to 1:1:1 K:Rb:Cs) show the largest suppression of \tc, consistent with the role of disorder.

\section{Conclusion}

In this work, we investigated alloys of the \AVS~kagome superconductors using both polycrystalline and single crystal samples. We identified that \KVS, \RVS, and \CVS~share a full solid-solution, with full miscibility on the alkali site. This provides a conceptual shift, wherein the \AVS~materials can be viewed as a continuum, with the three ``parent'' structures as the terminal ends of a single phase space. We presented the means to design \AVS~single crystal alloys of desired compositions, and explored the charge density wave (\tcdw) and superconducting temperatures (\tc) as a function of $A$-site composition. In contrast to the linear trends seen in the structural properties (e.g. cell volume), we observed continuous \textit{non-linear} changes in \tc~and \tcdw. For example, we identified a small dome of enhanced CDW stability along alloys of K$_{1-x}$Cs$_{x}$V$_3$Sb$_5$. By dramatically expanding the available chemical space, this work provides a new, materials-based route for probing the rich electronic phase diagram of the \AVS~kagome superconductors.

\section{Acknowledgments}

 This work was supported by the National Science Foundation (NSF) through Enabling Quantum Leap: Convergent Accelerated Discovery Foundries for Quantum Materials Science, Engineering and Information (Q-AMASE-i): Quantum Foundry at UC Santa Barbara (DMR-1906325). The research made use of the shared experimental facilities of the NSF Materials Research Science and Engineering Center at UC Santa Barbara (DMR- 1720256). The UC Santa Barbara MRSEC is a member of the Materials Research Facilities Network. (www.mrfn.org). 

\bibliography{AV3Sb5Alloy_v4}

\end{document}